\input epsf.tex
\documentclass[12pt]{article}
\usepackage{verbatim}
\usepackage{color}
\usepackage{graphicx}
\usepackage{amssymb,epsf,amsmath}
\csname @addtoreset\endcsname{equation}{section}

\def\bseq{\begin{subequation}}  % = 1a 1b
\def\eseq{\end{subequation}}
\def\Bar#1{\overline{#1}}                       % big bar

\newcommand{\beq}{\begin{equation}}
\newcommand{\eeq}{\end{equation}}
\newcommand{\bea}{\begin{eqnarray}}
\newcommand{\eea}{\end{eqnarray}}
\newcommand{\ena}{\end{eqnarray}}

\newcommand {\non}{\nonumber}

\renewcommand{\a}{\alpha}
\renewcommand{\b}{\beta}

\renewcommand{\d}{\delta}

\newcommand{\pa}{\partial}
\newcommand{\g}{\gamma}

\newcommand{\e}{\epsilon}

\renewcommand{\l}{\lambda}

\newcommand{\n}{\nu}

\newcommand{\Db}{\Bar{D}}

\newcommand{\Tr}{{\rm Tr}}

\newcommand{\intsup}{\int\!\! d^3xd^4\theta ~}

\begin{document}

\begin{titlepage}
{\hbox to\hsize{\hfill October 2009}}

\begin{center}
\vglue 0.99in
{\Large\bf INFRARED STABILITY OF 
\\ [.09in]
ABJ--LIKE THEORIES}
\\[.45in]
Marco S. Bianchi \footnote{marco.bianchi@mib.infn.it}, 
Silvia Penati\footnote{silvia.penati@mib.infn.it} ~and~
Massimo Siani\footnote{massimo.siani@mib.infn.it}\\
{\it Dipartimento di Fisica dell'Universit\`a degli studi di
Milano-Bicocca,\\
and INFN, Sezione di Milano-Bicocca, piazza della Scienza 3, I-20126 Milano,
Italy}\\[.8in]

{\bf ABSTRACT}\\[.0015in]
\end{center}

We consider marginal deformations of the superconformal ABJM/ABJ models which preserve ${\cal N}=2$
supersymmetry.
We determine perturbatively the spectrum of fixed points and study their infrared stability. 
We find a closed line of fixed points which is IR stable. 
The fixed point corresponding to the ABJM/ABJ models is stable under marginal deformations which 
respect the original $SU(2)_A \times SU(2)_B$ invariance, while deformations which break this group destabilize the theory 
which then flows to a less symmetric fixed point.  
We discuss the addition of flavor degrees of freedom. We prove that in general a flavor marginal superpotential 
does not destabilize the system in the IR. An exception is represented by a marginal coupling which mixes matter 
charged under different gauge sectors.  
Finally, we consider the case of relevant deformations which should 
drive the system to a strongly coupled IR fixed point recently investigated in arXiv:0909.2036 [hep-th].

${~~~}$ \newline
\vskip 10pt
\noindent
%PACS:
%03.70.+k, 11.15.-q, 11.10.-z, 11.30.Pb, 11.30.Rd (da sistemare (???))
%\\[.01in]

\end{titlepage}

\section{Introduction}

Recently, a renew interest in three dimensional superconformal Chern--Simons (CS) theories \cite{Schwarz}
has been triggered by the
formulation of the AdS4/CFT3 correspondence between CS matter theories and M/string theory. 
The low energy dynamics of a stack of $N$ M2--branes in M--theory probing a ${\cal C}^4/{\cal Z}_k$
singularity is given by a ${\cal N}=6$ supersymmetric two--level $(k,-k)$ CS theory for gauge group 
$U(N) \times U(N)$ with $SU(2)_A \times SU(2)_B$ invariant matter in the bifundamental 
representation \cite{ABJM}. In the decoupling limit $N \to \infty$, and for $k \to \infty$ with 
$\l \equiv N/k$ kept fixed, the CS matter theory is dual to a type IIA string theory on 
${\rm AdS}4 \times {\cal CP}^3$ background \cite{ABJM}.    
In the particular case of $N=2$ or for $k=1,2$  
supersymmetry gets enhanced to ${\cal N}=8$ \cite{ABJM, Gustavsson:2009pm}. For $N=2$ M--theory provides 
a dual description of the Bagger--Lambert--Gustavsson (BLG) model \cite{Bagger:2006sk}.

CS matter theories involved in the AdS4/CFT3 correspondence are of course at their superconformal fixed 
point.  
It is then natural to investigate the properties of these fixed points at quantum level in order
to establish whether they are isolated fixed points or they belong to a continuum surface of fixed points,
whether they are IR stable and which are the RG trajectories which intersect them. Since for $k \gg N$
the CS theory is weakly coupled, a perturbative approach is available.      

To this purpose we consider a ${\cal N}=2$ supersymmetric  $U(N)_k \times U(M)_{-k}$ CS theory 
with matter in the bifundamental representation, perturbated by
the most general superpotential compatible with ${\cal N}=2$ supersymmetry but generally breaking global 
nonabelian symmetries.
For particular values of the couplings our theory reduces to the ${\cal N}=6$ ABJM/ABJ superconformal theories 
\cite{ABJM, ABJ} (${\cal N}=8$ BLG theory \cite{Bagger:2006sk} for $N = M = 2$), while in general it describes marginal (but not exactly marginal) perturbations which drive the theory away from the superconformal point. 

At two loops, we compute the beta--functions for all the couplings and determine the spectrum of fixed points. 
We find an ellipse of ${\cal N}=2$ superconformal theories (thus confirming the results of 
\cite{Akerblom:2009gx}) which contains 
as a non--isolated fixed point the ${\cal N}=6$ ABJ/ABJM theories. At this point global symmetry 
gets enhanced to $SU(2)_A \times SU(2)_B$. The ellipse does not pass through the origin which is then an isolated solution. 
However, for $k$ sufficiently large compared to $N$ it passes arbitrarily close to the origin, so making 
the perturbative approach trustable.   

We study RG flows around the fixed points in order to investigate their IR stability.
The ${\cal N}=6$ ABJM/ABJ (or ${\cal N}=8$ BLG) theories turn out to be IR stable along the $SU(2)_A \times SU(2)_B$ 
preserving flow, whereas along any direction which breaks $SU(2)_A \times SU(2)_B$ global symmetry the theory flows to another
less symmetric superconformal model.  

More generally, for any fixed point on the ellipse there exists only one direction of stability which 
corresponds to the RG trajectory intersecting the ellipse in that point. 
Perturbations along any other direction will lead the theory to a different superconformal point. 
In that sense every single point is unstable under small perturbations. 

We may wonder whether our results change when restricting to the case of  marginal perturbations which 
respect some non-abelian global symmetry. To this end, we consider the subset of $SU(2)$ invariant marginal 
perturbations. 
We still find a closed line of fixed points describing $SU(2)$ invariant superconformal models. As in the more 
general case, while it is a line of IR stability each single point has only one direction of stability. 
The flow along any other direction leads the theory to a  different fixed point. In the particular case 
of ABJM/ABJ theories the direction of stability still corresponds to marginal perturbations which respect 
$SU(2)_A \times SU(2)_B$ invariance. 

We then study the effects of adding two sets of fundamental flavor degrees of freedom, one for each gauge group. 
In the presence of the most general marginal flavor perturbation we determine the surface of fixed points. It   
includes the point corresponding to 
the ABJ/ABJM models with flavors introduced in \cite{Gaiotto:2009tk,Hohenegger:2009as,Hikida:2009tp}.
Studying various examples of flavored theories, we find that in general the stability properties of the fixed points 
are not much different from the ones of the unflavored case. In fact, as long as we consider marginal couplings  
which do not mix matter in different gauge sectors, RG flows along the flavor directions always drive the system 
towards a fixed point in the IR. 

An exception is represented by a marginal perturbation which mixes matter charged under
$U(N)$ with matter charged under $U(M)$. In this case, at least at the perturbative order we are working, 
the corresponding coupling seems to introduce a direction of instability in the parameter space. 

Finally, we discuss the addition of perturbations which are relevant in the UV but 
which should drive the theory to a strongly coupled IR fixed point, as recently conjectured in
\cite{Martelli:2009ga}. In the perturbative regime we find that no other fixed point exists except 
the origin. This is an UV fixed point from which the theory escapes along RG flows when moving to
the low energy regime. This is then compatible with the existence of a strongly coupled IR point as the only 
point where superconformal invariance gets enhanced.

\section{The model and its $\b$--functions}

We choose to describe the ABJ/ABJM theories and their deformations in terms 
of an ordinary gauge theory with bifundamental matter \cite{VanRaamsdonk:2008ft}.   

In three dimensions, we consider a $U(N)_k \times U(M)_{-k}$ 
Chern--Simons theory for vector multiplets $(V,\hat{V})$ coupled to  chiral multiplets $A^i$ and 
$B_i$, $i=1,2$, in the $(N,\bar{M})$ and $(\bar{N},M)$ representations of the gauge group.
 
In ${\cal N}=2$ superspace the action reads (we use superspace conventions of \cite{superspace})
\bea  
{\cal S}
  &=& K \int d^3x\,d^4\theta \int_0^1 dt\: \Tr \Big[
  V \Db^\a \left( e^{-t V} D_\a e^{t V} \right) - 
  \hat{V} \Db^\a \left( e^{-t \hat{V}} D_\a
  e^{t \hat{V}} \right) \Big]   
  \non \\ 
  \non \\
   &+& \int d^3x\,d^4\theta\: \Tr \left( \bar{A}_i
  e^V A^i e^{- \hat{V}} + \bar{B}^i e^{\hat V} B_i
  e^{-V} \right) 
  \non \\
  \non \\
  &+& \int d^3x\,d^2\theta\:
    \left[  h_1 \, \Tr ( A^1 B_1 A^2 B_2) + h_2  \, \Tr (A^2 B_1 A^1 B_2)  \right] + {\rm h.c.}
   \label{action}
\eea
where $k=2\pi K$ is an integer which in the perturbative regime we take large ($\l \equiv N/k \ll 1$). 
The superpotential is  a classical 
marginal perturbation which respects ${\cal N}=2$ supersymmetry. For generic couplings the only global 
symmetries of the theory are the $U(1)_R$ R--symmetry and two $U(1)$'s acting as  
\bea
&& U(1)_1: \quad A^1 \rightarrow e^{i\a} A^1  \qquad ~, \qquad  U(1)_2: \quad B_1 \rightarrow e^{i\b} B_1
\non \\
&& \qquad \qquad A^2 \rightarrow e^{-i\a} A^2 \qquad , \qquad \qquad  \qquad ~ B_2 \rightarrow e^{-i\b} B_2
\eea
On the particular line $h_1 = - h_2$ in the space of the couplings the global symmetry gets enhanced to
$SU(2)_A \times SU(2)_B$. At the point $h_1 = -h_2 =1/K$ the symmetry $U(1)_R \times SU(2)_A \times SU(2)_B$
gets promoted to $SU(4)_R$ \cite{ABJM, Benna:2008zy} and we recover the ${\cal N}=6$ superconformal ABJ theory 
\cite{ABJ} and for $N=M$ the ABJM theory \cite{ABJM}.

\vskip 10pt
The theory can be quantized in a manifest ${\cal N}=2$ setup \cite{Avdeev:1992jt, us}. 
Perturbatively, UV divergences appear only at even orders in loops, so we 
concentrate on the renormalization of the theory at two loops. 
 
It is well known \cite{Avdeev:1992jt, KLL} that the gauge sector does not receive any UV correction. 
Moreover, a non--renormalization theorem for chiral integrals is still working in 3$d$ and allows to conclude 
that the superpotential does not get infinite corrections. Therefore, the only divergences that we 
need evaluate concern self--energy contributions to the matter sector.

\begin{figure}
  \center
  \includegraphics[width=0.8\textwidth]{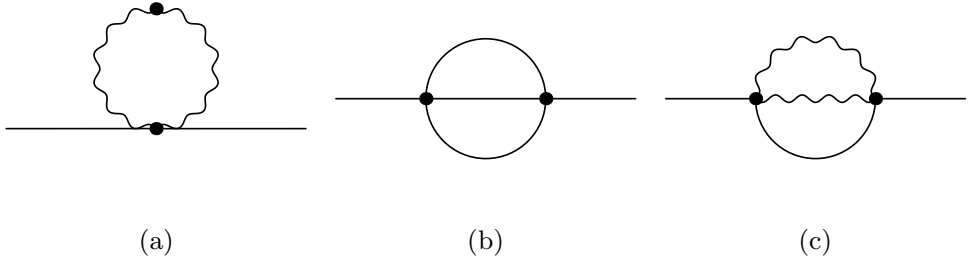}
  \caption{Two--loop diagrams contributing to the anomalous dimensions of fundamental chiral superfields. 
  Straight lines correspond to chirals, whereas wave lines to vectors.  }
\end{figure}

In dimensional regularization ($d=3-2\e$) and using ${\cal N}=2$ superspace techniques we compute 
two--loop divergences corresponding to the diagrams in Fig. 1. Renormalizing the theory according to
\bea
&&  A = Z_A^{-\frac12} A^{(0)} \qquad \qquad \qquad \; \; \bar A = 
  Z_A^{-\frac12} \bar A^{(0)}
\non \\
&&  B = Z_B^{-\frac12} B^{(0)} \qquad \qquad \qquad \; \; \bar B =
  Z_B^{-\frac12} \bar B^{(0)}
  \non \\
&&  h_j = \mu^{-2\e} Z_{h_j}^{-1} h_{j}^{(0)} \qquad \qquad \quad \bar h_j = \mu^{-2\e} Z_{\bar
  h_j}^{-1} \bar h_j^{(0)}
\eea
together with $K = \mu^{2\e}K^{(0)}$, we determine the anomalous dimensions, 
$\g =\frac12 \frac{\pa \log Z}{\pa \log \mu}$, of the fundamental fields. Leaving details for a 
future publication \cite{us}, here we quote only the result 
\bea
\g_{A} = \g_{B} = \frac{1}{32\pi^2} \Big[ 2 \, \frac{1 - NM}{K^2} 
  + (|h_1|^2+|h_2|^2) MN + (h_1 \bar h_2 + h_2 \bar h_1) \Big] \equiv \g
\label{gamma}
\eea 
The corresponding $\b$--functions are then given by
\beq
\b_{h_1} = 4 h_1  \g \qquad \qquad \b_{h_2} = 4 h_2  \g
\label{beta}
\eeq
as follows from the non--renormalization theorem for superpotentials.

\section{Fixed points and their IR stability}

We study the spectrum of fixed points and their stability in the case of real couplings.
We prefer to perform the rotation  
\beq 
y_1 \equiv h_1 + h_2 \qquad , \qquad y_2 \equiv h_1 - h_2
\label{y}
\eeq
and study RG flows in the $(y_1, y_2)$ plane.

In terms of these new couplings the superpotential in eq. (\ref{action}) can be rewritten as
\beq
\int d^3x\,d^2\theta\:
 \left[  \frac{y_1}{2} \Tr \,\Big( A^1 B_1 A^2 B_2  + A^2 B_1 A^1 B_2 \Big)
   + \frac{y_2}{4} \e_{ij} \e^{kl} \Tr \,(A^i B_ k A^j B_l) \right]
\eeq
It is then clear that the $y_2$--line (i.e. $y_1=0$) corresponds to the set of $SU(2)_A \times SU(2)_B$ invariant theories, 
whereas the $y_1$ coupling measures $SU(2)_A \times SU(2)_B$ symmetry breaking. 

We solve the equations $\frac{d y_i}{d t} = \b_{y_i}$ with $\b_{y_1} = 4 y_1 \g$, $\b_{y_2} = 4 y_2 \g$
and
\bea
\label{betay}
\g = \frac{1}{64\pi^2 K^2} \, \Big[ y_1^2 (MN+1) K^2 + y_2^2 (MN-1) K^2 - 4(MN-1) \Big]
\eea
The spectrum of nontrivial fixed points is given by the condition
\beq
y_1^2 (MN+1) + y_2^2 (MN-1) = \frac{4}{K^2} (MN-1) 
\label{fixedpts}
\eeq
which describes an ellipse in the space of the couplings (see Fig. 2), as already found in 
\cite{Akerblom:2009gx}. We note that in the large $M,N$ limit
and for $K \gg 1$ this becomes a circle with center in the origin and radius infinitesimally small. 
Therefore, the solutions fall inside the region of validity of the perturbative description. 
The particular point $(0, 2/K)$ corresponds to the ABJ/ABJM models.

The result (\ref{fixedpts}) read in terms of the original couplings $(h_1,h_2)$ 
states that in the class of scalar, dimension--two composite operators of the form
\beq
{\cal O} =  h_1 \, \Tr ( A^1 B_1 A^2 B_2) + h_2  \, \Tr (A^2 B_1 A^1 B_2)  
\eeq
there is only one exactly marginal operator.  This is the operator which allows the system to move along 
the fixed line.

\begin{figure}
  \center
  \includegraphics[width=0.5\textwidth]{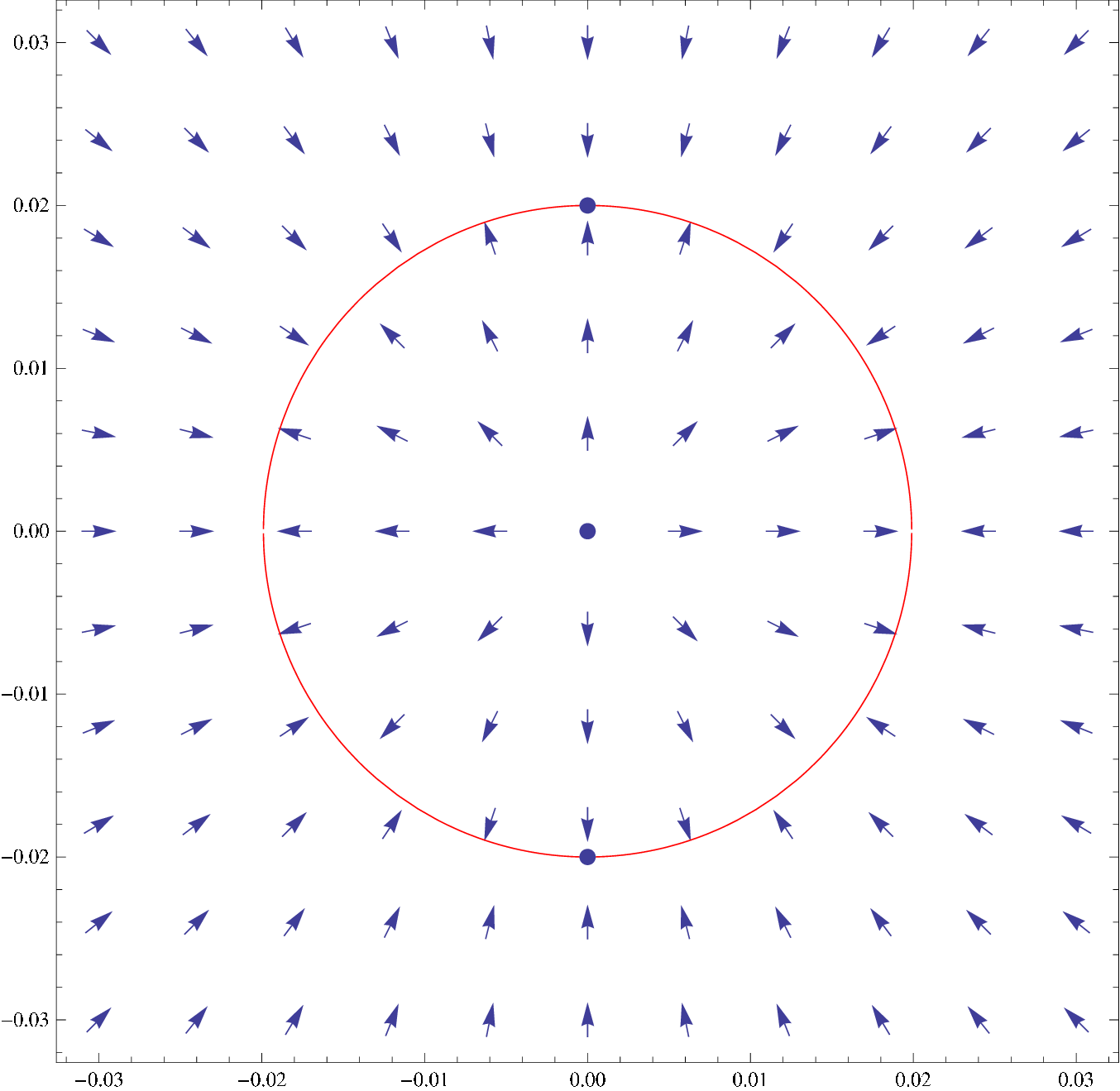}
  \caption{Line of fixed points and RG trajectories. The arrows indicate flows towards the IR. We have chosen $K=100$, $N=10$, $M=20$. }
\end{figure}

We now study the RG flows by solving the RG equations exactly. Using eq. (\ref{betay}) we can write 
\beq
\frac{d y_2}{d y_1} = \frac{y_2}{y_1}
\eeq
and the most general solution is $y_2 = C y_1$ with $C$ arbitrary. Therefore, in the $(y_1, y_2)$ plane
the RG trajectories are all the straight lines passing through the origin. 

Infrared flows can be easily determined by plotting the vector $(-\b_{y_1}, -\b_{y_2})$ 
in each point of the $(y_1,y_2)$ plane. The result is given in Fig. 2 where a number of interesting features 
arise. 

First of all the origin, corresponding to the free theory, is always an unstable point in the IR. 
Second, the line of fixed points is stable in the sense that the system always flows towards it. 
However, every single fixed point has only one direction of stability which corresponds to the RG trajectory 
passing through it. For any other direction of perturbation it is unstable since in general a small 
perturbation will drive the system to a different point on the line. 

In particular, the ABJ/ABJM fixed point is IR stable against 
small perturbations which respect the $SU(2)_A \times SU(2)_B$ symmetry (that is along the vertical line $y_1 = 0$), 
whereas if the perturbation breaks $SU(2)_A \times SU(2)_B$ the system will flow to a less symmetric fixed point.

\vskip 10pt
  
As we can argue from the previous results, at low energies the ABJ/ABJM system has a better response towards 
perturbations
which exhibit a certain amount of nontrivial global symmetries. Therefore, one might wonder whether restricting 
the class of marginal deformations by requiring non--abelian global symmetries to be preserved the previous
pattern could change. 

To answer this question we consider a class of perturbations which preserve a $SU(2)$ global symmetry out of
the $SU(2)_A \times SU(2)_B$ symmetry of the original ABJM/ABJ models. We replace the superpotential in (\ref{action}) with
\beq
\int d^3x\,d^2\theta\:
   \Tr  \left[  c_1 \Tr (A^i B_i)^2  + c_2 \Tr( B_i A^i)^2 \right] + {\rm h.c.}
\label{su2}
\eeq
where $c_1, c_2$ are two real constants.
In this case, a two--loop evaluation of the beta--functions leads to 
\beq
\b_{c_1} = 4 c_1  \g \qquad \qquad \b_{c_2} = 4 c_2  \g
\eeq
\beq
\g = \frac{1}{32\pi^2} \Big[ 2 \, \frac{1 - NM}{K^2} 
+ 4(c_1 + c_2)^2 (MN +1) + 4 (c_1^2+c_2^2) MN + 8 c_1 c_2  \Big]
\non
\eeq 
Proceeding as before, we determine the line of fixed points which in terms of  
the new variables $(y_1, y_2) = (2(c_1 + c_2), 2(c_1 - c_2))$ reads
\beq
3 y_1^2 (MN+1) + y_2^2 (MN-1) = \frac{4}{K^2} (MN-1) 
\eeq
In the space of the couplings this is still an ellipse similar to the one in Fig 2. 
The ABJM/ABJ models correspond to the point $(0, 2/K)$. 

The study of the IR flows is very similar to the previous case and leads to a behavior like the one drawn in 
Fig. 2. Therefore, we conclude that even in the presence of residual non--abelian global symmetries the 
only direction of stability for the ABJM/ABJ models is the $SU(2)_A \times SU(2)_B$ invariant line.

\section{Adding flavors}

We now discuss the addition of extra flavor degrees of freedom to the theory (\ref{action}) focusing on the 
effects that they have on the stability properties previously discussed.

We introduce flavor matter described by two couples of  
chiral superfields $Q_i, \tilde Q_i, \, i=1,2$ charged under the gauge groups and under a global
$U(N_f)_1 \times U(N_f^\prime)_2$. Precisely, $Q_1$ ($\tilde{Q}_1$) belongs to the 
(anti)fundamental of $U(N)$ while $Q_2$ ($\tilde{Q}_{2}$) belongs to the (anti)fundamental of $U(M)$.

We assign the following action 
\bea   
\label{action2}
S_{f} &=& \intsup \Tr \left( {\bar Q}^1 e^V Q_1 +
  \bar{\tilde{Q}}^{1} \tilde Q_{1} e^{-V} + \bar Q^2
  e^{\hat V} Q_2 + \bar{\tilde Q}^{2} \tilde Q_{2} e^{-\hat V} \right)
    \label{eqn:mat-action} \\ 
  \non \\  
  &&+ \int d^3x\,d^2\theta\:  \left[ \l_1 \Tr \, (Q_1 \tilde Q_1)^2 + \l_2 \Tr \,
  (Q_2 \tilde Q_2)^2 + \l_3 \Tr \, (Q_1 \tilde Q_1 Q_2 \tilde Q_2) \right.
   \non \\ 
  &&\left. \qquad \qquad \qquad \quad + \a_1 \Tr \, \left(\tilde Q_1 A^i B_i Q_1 \right)
  + \a_2 \Tr \, \left(\tilde Q_2 B_i A^i Q_2 \right) \right]  \, + \, {\rm h.c.}
  \non
\eea 
to be added to (\ref{action}).

Working out the two--loop beta--functions for this case, still redefining the bifundamental couplings as in 
(\ref{y}) we find (details will be given in \cite{us})
\bea 
\label{beta2}
&& \qquad \qquad ~~\b_{y_1} = 4y_1 \g  \qquad \qquad  \b_{y_2} = 4 y_2 \g 
\\ 
&& \b_{\l_1} = 4\l_1 \g_{Q_1}  \qquad \quad  \b_{\l_2} = 4 \l_2 \g_{Q_2}  
\quad \qquad \b_{\l_3} = 2\l_3 ( \g_{Q_1} + \g_{Q_2} ) 
\non \\
&& \qquad \b_{\a_1} = 2\a_1 ( \g +  \g_{Q_1} )  \qquad \quad   \b_{\a_2} = 2\a_2 ( \g + \g_{Q_2} )
\non
\eea
where
\bea
\label{gamma2}
 \g &=& \frac{1}{64\pi^2K^2} \Big[  y_1^2 (MN+1) K^2 + y_2^2 (MN-1) K^2 - 2(2MN + N N_f + M N_f^\prime -2) 
 \non \\ 
  &&~~~~~~~~~~~~~~~~ \qquad \quad + 2 |\a_1|^2 N N_f K^2  + 2 |\a_2|^2 M N_f^\prime K^2 \Big]
\non \\
\g_{Q^1} &=& \g_{\tilde{Q}_1} = \frac{1}{32\pi^2K^2} \Big[ -(2NM + N N_f + 1)
  \non \\
  &&~~~~~~~~~~~~~~~~ \qquad \quad  + 4|\l_1|^2\, (N N_f+1)K^2  + |\l_3|^2
  MN^\prime_f K^2 + 2|\a_1|^2  M N K^2  \Big] 
  \non \\
  \g_{Q^2} &=& \g_{\tilde{Q}_2} = \frac{1}{32\pi^2K^2 } \Big[ -(2NM + M N^\prime_f + 1)  
  \non \\
  &&~~~~~~~~~~~~~~~~ \qquad \quad  + 4|\l_2|^2\, (M N^\prime_f+1) K^2 + |\l_3|^2 N
  N_f K^2 + 2|\a_2|^2 M N K^2 \Big]  \non \\
\eea
In the space of the couplings the spectrum of fixed points describes a
four dimensional hypersurface given by the equations $\g = \g_{Q_1} = \g_{Q_2}  = 0$. A
particular point on this surface, corresponding to $h_1 = -h_2 = 1/K$,
$\l_1 = -\l_2 = 1/2K$, $\l_3=0$ and $\a_1 = -\a_2 = 1/K$, gives
the ABJ/ABJM models with flavors discussed in \cite{Gaiotto:2009tk,
Hohenegger:2009as,Hikida:2009tp}.

The IR behavior around a given fixed point $x_0$ can be inferred from the
study of the stability matrix ${\cal
M}_{ij} \equiv \frac{d \b_{\nu_i}}{d \n_j}(x_0)$, where $\n_j$ indicates a generic
coupling.  Diagonalizing this matrix, positive eigenvalues correspond
to stable directions, whereas negative eigenvalues signal IR
instability.

Due to the presence of a high number of couplings, the study of stability properties 
may become cumbersome. Therefore, we analyze particular cases separately.   

Choosing for simplicity $M=N$ (and then $N_f^\prime = N_f$), we begin by considering a theory with 
$\a_1 = \a_2 = 0$ in (\ref{action2}).
In this case, at the perturbative order we are working the set of RG equations for $(y_1,y_2)$  
decouples from the one for $(\l_1,\l_2,\l_3)$.  
Since the equations for $(y_1,y_2)$ are similar to the ones of the unflavored theory,  
the fixed points in the $y$--plane still describe an ellipse as in Fig. 2. In the $\l$--space, instead, 
they span the closed curve of Fig. 3.

\begin{figure}
  \center
  \includegraphics[width=0.3\textwidth]{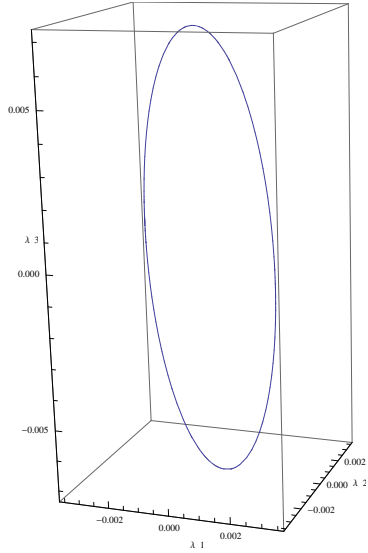}
  \caption{The ellipse of fixed points in the $(\l_1,\l_2,\l_3)$ space for the flavored model with $\a_i=0$. The parameters are $K=100$, $M=N=50$, $N_f=N_f\prime=5$. }
\end{figure}

The stability properties of the system on the $y$--plane do not get affected by the presence of flavor 
couplings, so they are analogous to the ones established for the unflavored theory. 
It is instead more interesting to investigate the IR stability 
of the fixed points when slowing moving along RG trajectories in the $\l$--space. 

To this end, we compute the stability matrix $\frac{\partial \b_{\l_i}}{\partial \l_j}$ at the 
ABJ--like fixed point given by
\bea
&& (y_1,y_2,\l_1,\l_2,\l_3) =
\non \\
&& \left(0 ,\frac{2}{K} \sqrt{ \frac{N^2 + NN_f -1}{N^2-1}}  , 
\frac{1}{2K} \sqrt{ \frac{ 2N^2 + NN_f +1}{NN_f+1}}, - \frac{1}{2K} \sqrt{ \frac{ 2N^2 + NN_f +1}{NN_f+1}}, 0\right)
\non \\
\eea
We find the diagonal matrix 
\bea
\frac{\partial \b_{\l_i}}{\partial \l_j}\Big| = diag \left( \frac{N (2 N+{N_f})+1}{4 K^2 \pi
 ^2}, \frac{N (2 N+{N_f})+1}{4 K^2 \pi ^2}, 0 \right)
\eea
The positive eigenvalues state that the system is IR stable along the $\l_1$ and $\l_2$ flows. In the $\l_3$ direction we find 
a vanishing eigenvalue, probably an artifact of the perturbative order we are working at. 
In order to solve the corresponding
indetermination we slightly move away from the fixed point along the $\l_2$ flow and evaluate the 
stability matrix for $\l_2 \to \l_2 + \d $.  We still find a diagonal matrix
\bea
\frac{\partial \b_{\l_i}}{\partial \l_j}\Big|_{\l_2 + \d} = 
diag \left( \frac{N (2 N+{N_f})+1}{4 K^2 \pi^2}, \frac{N (2 N+ N_f) + 1}{4 K^2 \pi^2}, - \frac{\delta (N {N_f}+1) 
\sqrt{\frac{2 N^2}{N {N_f}+1}+1}}{4 K \pi ^2} \right) \non \\
\eea
where $\d$--corrections to the first two eigenvalues have been neglected. We note that for $\d > 0$ 
the third entry of the matrix is negative and the $\l_3$--flow in that region is IR unstable. 
We conclude that $\l_3$--perturbations, the ones which mix the two flavor sectors, destabilize the system 
in the IR. 

We now consider a theory with $\l_i=0$ for any $i=1,2,3$ and $\a_1 = - \a_2 \equiv \a \neq 0$ in (\ref{action2}). 
The effect of turning on 
couplings between fundamental and bifundamental fields is to render the system of RG equations nontrivially coupled
(see eqs. (\ref{beta2}, \ref{gamma2})) already at two loops. The study of RG flows has then to be performed in the three dimensional space $(y_1,y_2,\a)$. 
Setting the beta--functions to zero, besides the trivial fixed point corresponding to the free theory, 
we find three ellipses of fixed points belonging to three bidimensional planes orthogonal to 
the $\a$ direction and placed at  $\a =  \pm \frac{1}{K}\sqrt{1+ (NN_f+1)/2N^2}$ and $\a = 0$, respectively.  

We can evaluate the beta functions in each point of the parameter space and 
find the RG trajectories. On the plane $y_1 = 0$ they look as in Fig. 4 where in red we have indicated the particular fixed 
points corresponding to the intersections of the ellipses with the $y_1=0$ plane. 

Once again, we find that the free theory is reached in the UV. Concerning interacting theories, from Fig. 4 
we see that flavored models without flavor superpotential ($\a =0$) are always IR unstable, whereas stability is reached 
with a nontrivial superpotential at $\a = \pm \frac{1}{K}\sqrt{1+ (NN_f+1)/2N^2}$. 
In particular, it is easy to see that there are RG trajectories which  interpolate between the UV 
ellipse $\a=0$ and the IR ellipses.

In Fig. 4 we have chosen to project on the $y_1=0$ plane for simplicity, but there is nothing special 
about this plane. The same configuration of RG trajectories is obtained by projecting on any other 
plane of the form $y_1 = m y_2$.   
 
 \begin{figure}
  \center
  \includegraphics[width=0.65\textwidth]{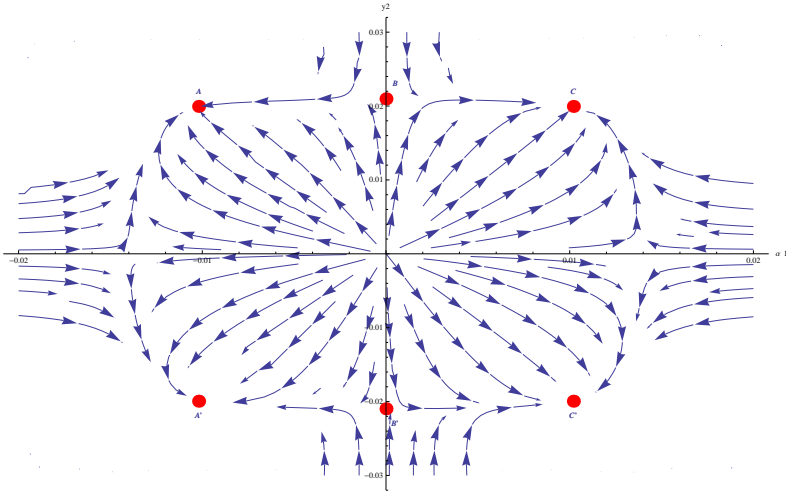}
  \caption{RG trajectories for the flavored model with $\l_i=0$ and $\a_1=\a_2 \neq 0$. The arrows indicate flows towards the IR. We have chosen $K=100$, $M=N=50$, $N_f=N_f\prime=5$. }
\end{figure}
 
Similar patterns arise when adding flavor matter to $SU(2)$ invariant theories (\ref{su2}) (note that
the flavor superpotential in (\ref{action2}) respects the global symmetries of the bifundamental
sector).

\section{A relevant perturbation}

As a concluding remark, we consider ${\cal N}=2$ CS--matter theories with two extra propagating
chiral superfields in the adjoint and a superpotential given by
\beq
\int d^3x\,d^2\theta\:
 \left[  s \, \Tr ( \Phi_1^3)  +  s \, \Tr ( \Phi_2^3) + t \, \Tr (B_i \Phi_1 A^i) +  t \, 
 \Tr (A^i \Phi_2 B_i) \right] + {\rm h.c.}
\label{pot2} 
\eeq
This kind of theories should flow to a strongly coupled fixed point in the IR since, as conjectured in
\cite{Martelli:2009ga}, they have a dual description in terms of a ${\rm AdS}4 \times V_{5,2}/{\cal Z}_k$
supergravity solution. 

In the UV region (\ref{pot2}) is a relevant perturbation with dimension--$\frac12$ couplings.
A perturbative evaluation of the beta--functions requires computing the two--loop diagrams of Figs. 1a),
1c) with $(\Phi_i, A^i, B_i)$ as external fields.  Setting $N=M$ for simplicity, in the large $N$ limit 
the result is 
\bea
&& \b_s = 3 s \, \g_{\Phi} \qquad  \qquad \qquad \quad \quad \g_{\Phi} = -\frac{1}{8\pi^2} \frac{N^2}{K^2} 
\\
&& \b_t = t \, ( \g_{\Phi} + \g_A + \g_B) \qquad  \quad ~\g_A = \g_B = -\frac{1}{16\pi^2} \frac{N^2}{K^2}
\non
\label{betarel}
\eea 
The only perturbatively accessible fixed point is $s=t=0$ which, according to the 
sign of the beta--functions, is reached at high energies. 
Therefore, the theory is free in the UV but naturally flows to a strongly coupled system in the IR. 
Such a behavior is similar to what we have found 
for the ABJ--like theories. However, in contrast with the previous case where under suitable requirements
on the gauge coupling the IR fixed points are visible perturbatively, for the present theory they are not
and other methods need be used to establish the existence of superconformal points \cite{Martelli:2009ga}. 

We note that our conclusions are not an artifact of the two--loop approximation. 
In fact, by dimensional analysis it is easy to realize that there are no contributions to the beta--functions
proportional to the chiral couplings $(s,t)$ at higher orders, being the gauge corrections the only possible 
sources of divergences. Therefore, no extra fixed points other than the free theory can be found 
perturbatively. This is an obvious consequence of supersymmetry and of the dimensionful nature of the 
chiral couplings. Even the addition of a SYM term in the original action which would not be excluded by 
the IR results of \cite{Martelli:2009ga} cannot change the analysis since in Feynman diagrams the replacement of 
CS vertices with SYM vertices improves the convergence of the integrals.

\vskip 25pt
\section*{Acknowledgements}
\noindent 

This work has been supported in part by INFN and MIUR-PRIN prot.20075ATT78-002.

\end{document}